\documentclass[10pt,reprint]{iopart}

\usepackage{graphicx}
\usepackage{psfrag}
\usepackage{color}
\usepackage{amsbsy}
\usepackage{amssymb}
\usepackage{iopams}  

\def\a{\alpha}
\def\b{\beta}

\def\d{\delta}
\def\ve{\varepsilon}

\def\t{\theta}

\begin{document}

  \title{Steady-state properties of a totally asymmetric exclusion process 
         with particles of arbitrary size}

  \title[]{}

  \author{Gregory W. Lakatos and Tom Chou\dag}
  \address{Institute for Pure and Applied Mathematics, UCLA, Los Angeles, CA, 90095, USA}

\begin{abstract}

The steady-state currents and densities of a one-dimensional totally asymmetric exclusion process
(TASEP) with particles that occlude an integer number ($d$) of lattice sites are computed using various
mean field approximations and Monte Carlo simulations.  TASEP's featuring particles of arbitrary size are
relevant for modeling systems such as mRNA translation, vesicle locomotion along microtubules, and
protein sliding along DNA.  We conjecture that the nonequilibrium steady-state properties separate into
low density, high density, an maximal current phases similar to those of the standard ($d=1$) TASEP.  A
simple mean field approximation for steady-state particle currents and densities is found to be
inaccurate. However, we find {\it local equilibrium} particle distributions derived from a discrete Tonks
gas partition function yield apparently exact currents within the maximal current phase. For the
boundary-limited phases, the equilibrium Tonks gas distribution cannot be used to predict currents, 
phase boundaries, or the order of the phase transitions.  However, we employ a refined mean field
approach to find apparently exact expressions for the steady state currents, boundary densities,
and phase diagrams of the $d\geq 1$ TASEP.  Extensive Monte Carlo simulations are performed to
support our analytic, mean field results.

\end{abstract}

\ead{tomchou@ucla.edu}



\section{Introduction} 

There has been much recent interest in Asymmetric Exclusion Processes (ASEPs), with both
periodic and open boundaries \cite{DER92,DER93,EVANS,DER98}.  In an exclusion process,
particles move on a one dimensional lattice.  With each movement a particle may move one
lattice site to the right with rate $p$, or to the left with rate $q$.  A particle may move to a
lattice site only if that site is not already occupied by another particle.  For a lattice with open
boundaries, particles are injected into the chain with rate $\a$ only if the first (leftmost) site is
empty. A particle occupying the last site is extracted with rate $\b$.  Similarly, if reverse steps
are allowed (partially asymmetric), injection into the rightmost lattice site occurs with rate $\d$
while extraction from the leftmost lattice site happens with rate $\gamma$.  The steady-state
currents, particle densities, and correlations of the ASEP have been studied extensively using
both exact matrix product methods and mean field approximations
\cite{DER92,DER93,DOMANY,RITT}. The exact steady-state currents in the infinitely long
lattice limit are summarized by a diagram depicting three regions; the maximal current, entry
limited current, and exit limited current phases.  Additional sub-phases corresponding to
different density {\it profiles}, or domain walls, arise within the exit and entry limited phase
regions \cite{KOLO2,NAGY}.

ASEPs have been used to gain insight into the nonequilibrium statistical mechanics arising in
practical settings such as traffic flow \cite{TRAFFIC,TRAFFIC2}, ion channels
\cite{CHOU99,PROTON}, mRNA translation \cite{MRNA1,MRNA2}, and vesicle translocation
along microtubules \cite{FREY}.  The standard ASEP assumes fixed transition rates for
particles whose step sizes equal their diameters.  However, in numerous applications, the
driven particles and the underlying length scales of the microscopic  transitions (steps) are
not identical; {\it i.e.}, the particle may be large and occlude $d > 1$ lattice sites. Specific
biophysical examples include ribosomes moving along mRNA and large molecules or vesicles
that are shuttled by motor proteins along microtubules.  Ribosomes are $\sim 20$nm in
diameter, but move by $\sim 2$nm steps, codon by codon.  Motor proteins that carry vesicles
($\gtrsim 50$nm) typically move with $\sim 5$nm steps along microtubules.  For these
applications $d \approx 10$.

Steady-state properties of the $d = 1$ ASEP (such as currents, densities, and their
correlations) have been computed exactly using matrix product and recursion methods
\cite{DOMANY}.  When the size of the particles is greater than the size of a single lattice site,
the matrix product method is inapplicable; however, diffusion constants and particle velocities
have been calculated for two particles of arbitrary size \cite{WADATI}.  Dynamical exponents
have also been derived using Bethe {\it ansatz} solutions \cite{ALCARAZ}. In this paper, we
use mean field analyses to derive apparently exact, simple formulae for steady state currents
and densities of a {\it totally} asymmetric ($q=0$) exclusion process (TASEP) with $d\geq 1$.
Results for partially asymmetric processes, where particles can move in both directions, are
straightforwardly derived from similar methods and will not be presented here. For the
TASEP, we begin with the {\it ansatz} that the qualitative features of the current phase
diagram remain intact when $d > 1$.  Namely, that the steady-state currents exist in one of
three regimes, high density (with currents determined by the exit rate), low density (with
currents determined by the particle entrance rate), and the maximal current phase (with
current determined by the interior hopping rate).

In the following section, we  consider a simple mean field theory that does not explicitly take
into account the infinitely repulsive interactions of the large particles. The discrepancy
between these qualitative results and those derived from Monte Carlo simulations are evident.
We then derive quantitatively accurate analytical approaches, based on discrete Tonks gas
partition functions and explicit state enumeration near the lattice boundaries.  Results from
both simple mean field, Tonks gas distribution, and the refined mean field approaches are
compared  with Monte-Carlo (MC) simulations in Section 3. We summarize and discuss
extensions of our approach to partially asymmetric exclusions in Section 4.


\section{Mean Field Theories}

Consider identical particles of diameter $d \geq 1$ (in units of lattice sites) that are driven
through a one-dimensional lattice having $N \gg d$ sites. The particles interact through hard
core repulsion. The ratios of entrance and exit rates, and steady state currents to the rate
$p$ are denoted $\a$, $\b$, and $J$, respectively. With each motion, the particles move one
lattice site to the right.

\subsection{Simple Mean Field}

The simplest mean field approximation to the steady-states of this system assumes that the
particles are on average uniformly distributed along the lattice without anti-correlations
arising from their hard core repulsive interactions.  The probability $P_{i}$ that a particle is at
site $i$ is assumed to be independent of $i$. In this discussion, a particle's position in the
lattice is determined by the location of its left edge.  The left edge of a particle of size $d$ can
move forward only if the $d$ sites ahead of it are unoccupied.  Setting $p$  equal to one
(rescaling all rates in units of $p$), the steady state current is then:

  \begin{equation}
     J = P_{i}(1-P_{i+1})(1-P_{i+2})\cdots(1-P_{i+d}).
  \end{equation}

\noindent Away from boundaries, the locally uniform probabilities $P_{i} \approx \rho$ give 


  \begin{equation}
     J = \rho (1-\rho )^{d}.
  \end{equation}

\noindent The maximum current that this uniform state can sustain, and its associated density
is found by setting $\partial J/\partial \rho = 0$, which yields


  \begin{equation}
     \rho_{max}= {1\over (d+1)}, \quad 
    J_{max} = { d^{d} \over (d+1)^{d+1}}.
  \end{equation}

\begin{figure}
    \begin{center}
      \includegraphics[height=2in]{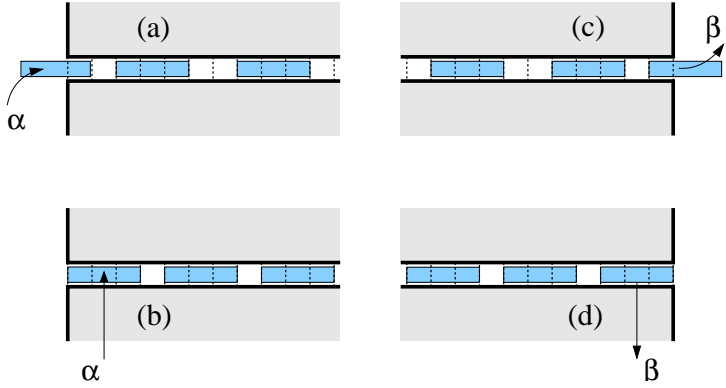}
    \end{center}
    \caption{Entry and exit mechanisms. (a) Incremental entry. (b) Complete entry 
      into the first $d$ sites. (c) Incremental exit. (d) Complete, one-step exit as the 
      particle reaches the last site.} 
    \label{ENTRY}
  \end{figure}

Now consider particle entry at the left end of the lattice, or particle exit from the right end. 
Figure \ref{ENTRY} shows two extreme limits for the rules of entry and exit when $d>1$. For
example, a particle may only enter incrementally, one lattice site at a time (Fig. 
\ref{ENTRY}a), or it may enter completely in one step when the first $d$ sites are empty
(Fig. \ref{ENTRY}b). One can also easily define an intermediate injection scheme
appropriate for the specific application at hand. For example, in mRNA translation, the entry
site for ribosomes must be completely cleared before another ribosome can bind.  Similar
microscopic considerations also permit different exit schemes (Figs.  \ref{ENTRY}c,d).
However, in the $N\rightarrow \infty$ limit, it is easy to see that the steady-state current and
densities are independent of the different entrance/exit schemes.  For example, partial entry
is equivalent to complete entry into a lattice that is extended by $d-1$ sites at the entrance
boundary, while partial exit is equivalent to complete, immediate exit of a lattice extended by
$d-1$ sites at the right boundary. Since $d/N \rightarrow 0$, the differences arising from the
variations in entrance and exit are finite-sized effects that vanish in the macroscopic limit. For
the sake of simplicity and consistency, we will assume complete entry and exit (Figs.
\ref{ENTRY}b,d) in all of our subsequent work.

For the entry-limited current, we set $\a\prod_{j=1}^{d} (1-P_{j}) =
P_{1}\prod_{j=1}^{d}(1-P_{1+j})$ and approximate $P_{i+j}\approx P_{1}$.  The  occupation
probability near the left entry site is thus $P_{1}\approx \a$ and the entry-limited steady-state
current becomes

  \begin{equation}
    J_{L}  = \a(1-\a)^{d},
    \label{JL0}
  \end{equation}

\noindent which we obtain if $\b$ is large enough (so that extraction is not the rate-limiting
step) and $\a(1-\a)^{d} < {d^{d}\over (d+1)^{d+1}}$, or equivalently, when $\a \leq 1/(d+1)$. 
Similarly, when the current is exit rate limited, $\b P_{N-d+1}\prod_{j=1}^{d-1}(1-P_{N-d+1+j})
= P_{N-d}\prod_{j=1}^{d}(1-P_{N-d+j})$. For slowly varying occupation probabilities near the
end of the chain, $P_{N} \approx 1-\b$,  and 

  \begin{equation}
    J_{R} = \b^{d}(1-\b).
    \label{JR0}
  \end{equation}

\noindent The extraction rate limited current (\ref{JR0}) holds when $\b \leq d/(d+1)$ and $\a
\geq 1/(d+1)$. This simple mean field theory ignores the {\it additional} exclusion between
particles with sizes greater than a single lattice site.  The resulting steady-state currents and
densities are described by the following regimes:

  \begin{equation}
    \begin{array}{rlll}
      \fl \mbox{(I)} & \displaystyle \a < {1\over d+1},\, \b^{d}(1-\b)\geq \a(1-\a)^{d} & \:\, 
      \displaystyle J_{L} = \a(1-\a)^{d} & \:\, 
      \rho_{1} = \a \\[13pt]
      \fl \mbox{(II)} & \displaystyle \b < {d \over d+1},\, \a(1-\a)^{d} \geq  \b^{d}(1-\b) 
      & \:\, \displaystyle 
      J_{R} = \b^{d}(1-\b) & \:\,
      \rho_{N}  = 1-\b  \\[13pt]
      \fl \mbox{(III)} & \displaystyle \a \geq {1\over d+1},\, \b \geq {d \over d+1} 
      &\:\, \displaystyle J_{max}= {d^{d} \over (d+1)^{d+1}} & \:\,
      \displaystyle \rho_{{N\over 2}} = {1\over d(d+1)}.
    \end{array}
    \label{BOUNDARIES0}
  \end{equation} 

The currents, boundary densities, and interior densities (\ref{BOUNDARIES0}) reproduce the
known results for the $d=1$ TASEP.  However, we do not expect these mean field results to
produce accurate density profiles \cite{DER93,DER98}. Moreover, the boundary densities are
expected to be inaccurate since they are independent of $d$. The predicted currents
(\ref{BOUNDARIES0}) are continuous across the $\a-\b$ parameter space and the transitions
are first order between phases (I) and (II) and second order between all other phases with a
$d$-dependent discontinuity 

  \begin{equation} 
\fl \quad \mbox{as}\, \ve \rightarrow 0, \,\,\,   \displaystyle {\partial^{2} J\over \partial
      \a^{2}}\bigg|_{\a=\a^{*}+\ve}\!\!\!\!\!- {\partial^{2} J\over \partial
      \a^{2}}\bigg|_{\a=\a^{*}-\ve} \!\!\!\!\! = {\partial^{2} J\over \partial
      \b^{2}}\bigg|_{\b=\b^{*}+\ve}\!\!\!\!\! - {\partial^{2} J\over \partial
      \b^{2}}\bigg|_{\b=\b^{*}-\ve}\!\!\!\!\! = {d^{d-1}\over (d+1)^{d-2}}.
    \label{JUMP0}
  \end{equation}

\noindent The currents, densities, and phase boundaries will now be compared with more accurate
mean field theories and numerical simulations.

\subsection{Tonks Equilibrium Distribution} 

We shall see that the simple mean field approximation developed in the previous section is
inaccurate for $d > 1$.  A more refined approximation for a TASEP with locally uniform
particle distributions is to consider the Tonks gas in {\it equilibrium}. For example, in an
infinite one-dimensional lattice at maximal current in the steady-state, the density distribution
is uniform over the lattice except near the ends.

A fixed section of length $N$ in the interior of the chain that contains $n \leq N/d$ uniformly
distributed particles can be treated as an {\it equilibrium} ensemble.  This is because in an
infinitely long chain, we can translate the segment $N$ at the mean particle velocity.  This
frame of reference allows us to treat the configurations sampled by the particle within $N$ as
that arising from the {\it equilibrium} grand canonical ensemble.  In the large $n$ limit, we can
use the canonical partition function of a gas of hard rods on a discrete lattice (discrete Tonks
gas) to compute the likelihood of configurations that support and contribute to the net
particle current.

\begin{figure}
    \begin{center}
      \includegraphics[height=1.8in]{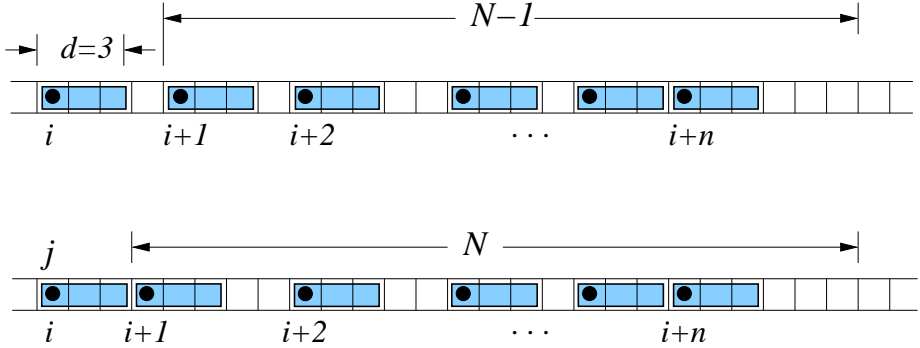}
    \end{center}
    \caption{(a) Constraints and a configuration associated with the 
      calculation of $Z(n,N)$. (b) Typical configuration for the calculation of $Z(n,N-1)$.}
    \label{TONKS}
  \end{figure}

Consider the case where the $i^{th}$ particle from the entrance of the lattice has its left edge
located at site $j \,\,(x_{i}=j)$.  The probability that lattice site $j+d$ is empty (in other words,
that the $i^{th}$ particle is free to move to the right) is calculated from the ratio $P(x_{i+1}>
j+d\vert x_{i} = j) = Z(n,N-1)/Z(n,N)$, since $Z(n,N-1)$ corresponds to the number of states of
$n$ particles where particle $i+1$ occupies sites to the right of $j+d$.  One specific
configuration in this sum is shown in Fig.  \ref{TONKS}a.

The partition function for $n$ particles of length $d$ confined to a length of $N\geq nd$
lattice sites is \cite{LATTICEZ}

  \begin{equation}
    Z(n,N) = \left(\begin{array}{c} N-(d-1)n \\[13pt]
      n\end{array}\right),
      \label{Z}
  \end{equation}

\noindent which leads to 

  \begin{equation}
    \begin{array}{rl}
      P(x_{i+1}>j+d\vert x_{i}=j) & \displaystyle = {Z(n,N-1)\over Z(n,N)} \\[13pt]
      \: & \displaystyle = {N-(d-1)n-n \over N-(d-1)n}  = {1-\rho d \over 1-\rho(d-1)},
    \end{array}
  \end{equation}

\noindent where $\rho \equiv n/N$. The steady-state current in a uniform phase is thus

  \begin{equation}
    J = P(x_i = j)P(x_{i+1}>j+d\vert x_{i}=j)
    = \rho {1-\rho d\over 1-\rho (d-1)}.
    \label{JUNIFORM}
  \end{equation}

\noindent The maximal current $J_{max}$ and its associated density $\rho_{max}$ is found by
setting $\partial_{\rho} J = 0$, which yields

  \begin{equation}
    \rho_{max} = {1\over \sqrt{d}(\sqrt{d}+1)}  \quad \mbox{and}\quad
    J_{max} = {1 \over (\sqrt{d}+1)^{2}}. 
    \label{JMAX}
  \end{equation}

Equation (\ref{JMAX}) reduces to the standard TASEP result \cite{DER92,DER93} when
$d=1$  and $J_{max}, \rho_{max} \sim 1/d$ as $d\rightarrow \infty$. The steady-state current
will be the maximal current $J_{max}$ when the ASEP is forced under periodic boundary
conditions (a ring) or as long as the injection and extraction rates of an open boundary ASEP
are large enough for the internal moves to be the overall rate limiting steps.

The Tonks gas approach would not be expected to yield accurate results in the boundary
limited current regimes (I) and (II) since these currents are determined by the behavior at the
boundaries where densities are not uniform. Even near the boundary limited and maximum
current phase boundaries, when the density profile varies slowly (as power laws)
\cite{DER93,DER98,DOMANY}, one cannot assume uniform density and apply a local (near
the entrance and exits) Tonks gas.    Nevertheless, within the maximal current phase, currents,
and densities derived from the Tonks gas approach compare extremely well with those found
from Monte Carlo simulations.

\subsection{Refined Mean Field Theory for Boundary-limited Currents} 

Here, we develop a mean field theory that accurately models the behavior of the particles at
the ends of the lattice, and matches with the Tonks gas theory as the injection $\a$ and
extraction $\b$ rates are varied across phase boundaries. The resulting steady-state currents
are also found to agree, within error, with results from Monte Carlo simulations.

\begin{figure}
    \begin{center}
      \includegraphics[height=2.0in]{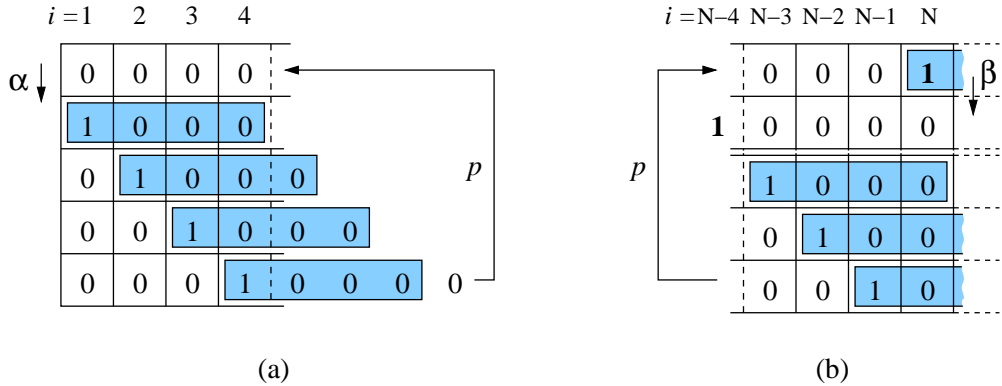}
    \end{center}
    \caption{Mean field scheme for particles of size $d=4$ near boundaries. 
  (a) $d+1$ distinct states at the entrance 
      region. (b) $d+1$ distinct states at the exit region. All the states that are within $d$ 
      microscopic steps of the first state pictured are included.}
    \label{STATES}
  \end{figure}

First consider the entrance site at the left of the chain as shown in Fig. \ref{STATES}a. 
Upon balancing the probability currents into the first state  shown in Fig. \ref{STATES}a, 

  \begin{equation}
    J_{L} = \a P(x_{1}>d)= P(x_{1}=d, x_{2}>2d).
    \label{JL1}
  \end{equation}

\noindent Although the simple mean field assumption used to derive $J_{L}$ and $J_{R}$
(Eqs. (\ref{JL0}) and (\ref{JR0})) completely ignore exclusion beyond one lattice site, relative
probabilities can be used to derive a refined mean field theory for the boundary limited
currents.  The exclusive nature of the particles are partially incorporated in this instance by
explicitly enumerating  the $d+1$ allowed  states. For example, the probability that the first
$d$ sites are unoccupied is 

  \begin{equation}
    P(x_{1}>d) = {(1-\t)^{d} \over d\t(1-\t)^{d-1} + (1-\t)^{d}},
  \end{equation}

\noindent where $\t$ is the relative likelihood that a site is occupied assuming a
nearly uniform particle density.  The  normalisation factor $d\t(1-\t)^{d-1} + (1-\t)^{d}$ is the
sum of the weights of all possible states of the first $d$ lattice sites (Fig. \ref{STATES}a).  The
probability that the first particle is at site $d$, and is not blocked from moving one lattice site
to the right is:

  \begin{equation}
  P(x_{1}=d, x_{2}>2d) = 
  {\t(1-\t)^{d} \over d\t(1-\t)^{d-1} + (1-\t)^{d}}.
  \end{equation}

\noindent We have assumed a slowly varying density so that $\t$ represents a mean over the first $d$
sites.    The mean occupation of the first $d$ sites is then

  \begin{equation}
  \rho_{L} \equiv \sum_{i=1}^{d}\rho_{i}= {\t(1-\t)^{d-1} \over d\t(1-\t)^{d-1}+(1-\t)^{d}} = {\t \over 1+(d-1)\t}.
  \label{RHO1}
  \end{equation}

\noindent Upon using Eq. (\ref{JL1}), $\t = \a$, and 

  \begin{equation}
  \rho_{L} = {\a \over 1+(d-1)\a}, \, J_{L} = {\a (1-\a) \over 1 + (d-1)\a}.
  \label{JL2}
  \end{equation}

For the exit rate-limited (small $\b$) steady-state current, we consider  all possible states
between  sites $N-d+1$ to $N$.  Here, since the last particle lingers at site $N$, the
occupation fractions will not be slowly varying over the last $d$ sites (cf. Fig. 
\ref{PROFILE}c).  For small exit rates, we expect that close-packing of the last few particles
near the end of the chain will split the densities into two branches.  One branch corresponds
to a lingering close-packed state with enhanced occupation fractions $\t_{i}$ at
$i=N,N-d,N-2d,\ldots$.  These densities will be slowly varying allowing us to make the
approximation $\t_{N}\approx \t_{N-d}\approx \t_{N-2d} \ldots \equiv \t_{N}$.  Sites belonging
to the second density branch lie between the high occupancy sites. These second-branch
sites tend to be occupied when the particles are not close-packed; for example, shortly after a
particle is extracted from the lattice.  These densities, $\t_{N-1}, \t_{N-2}, \ldots, \t_{N-d+1}
\equiv \t$ are smaller than $\t_{N}$ by a factor of approximately $\beta$.  

Transitions between states corresponding to occupancy in these two sets of sites are
considered.  Conservation of the probability of the first state (only site $N$ occupied)
involves the flux $\b$ out of that state, and the replenishment from the last state occurring at
rate $p$.  The steady-state current can be found from

  \begin{equation}
    J_{R} = \b P(x_{n}=N) = P(x_{n}= N-1).
    \label{JR1}
  \end{equation}

\noindent  Since in each ``unit cell'' there is only one high occupancy type site,
(site $N$), there is no additional normalization factor
and the probability that the last particle is situated at site $N$, ready to exit with rate
$\b$ is simply $P(x_{n}=N) = \t_{N}$. The probability that the last particle is at $N-1$ is

\begin{equation}
P(x_{n}=N-1) = {\t (1-\t)^{d-2} (1-\t_{N}) \over (1-\t)^{d-1}+ (d-1)\t(1-\t)^{d-2}},
\end{equation}

\noindent where $(1-\t)^{d-1}+ (d-1)\t(1-\t)^{d-2}$ is the 
total weight for all states of the sites $N-1, N-2, \ldots, N-d+1$
(the last four rows in Fig. \ref{STATES}b). Upon solving (\ref{JR1}), we 
find 

\begin{equation}
{\b\t_{N} \over 1-\t_{N}} = {\t\over 1+(d-2)\t}. 
\label{BETATN}
\end{equation}

\noindent To find another relationship between $\t$ and $\t_{N}$, equate the flux into and out
of the state where site $N-3$ is occupied: 

\begin{equation}
\t_{N}(1-\t_{N})(1-\t)^{d-1} = \t(1-\t)^{d-2}(1-\t_{N}).
\label{PPTN}
\end{equation}

\noindent  Upon solving equations (\ref{BETATN}) and (\ref{PPTN}), we 
find

\begin{equation}
   \t_{N} \equiv  \rho_{R}^{+} = {1-\b \over 1+(d-1)\b},\, J_{R} = {\b (1-\b) \over 1 + (d-1)\b},
    \label{JR2}
\end{equation}

\noindent and mean occupations at the low density sites $\rho_{N-1} =\rho_{N-2},
\ldots,\rho_{N-d+1} \equiv \rho_{R}^{-}= \b\rho_{R}^{+}=J_{R}$.  

The currents and densities (\ref{JL2}) and (\ref{JR2}) reduce to the exact results of the
standard TASEP for $d=1$. Unlike in the $d=1$ case, particle-hole symmetry is no longer
preserved for particles that occlude more than a single lattice site.  A different state
enumeration near the two ends of the lattice must be considered, but the resulting
expressions for the currents $J_{L}$ and $J_{R}$ remain symmetric with interchange of
$\a\leftrightarrow \b$.  

\subsection{Matching mean field phase boundaries} 

The different approximations for the steady-state currents and densities cover all allowable
values of the parameters  $\a,\b$.  Equating the various currents (\ref{JMAX}), (\ref{JL2}), and
(\ref{JR2}), we find the following phase boundaries between the different current regimes
 
  \begin{equation}
    \begin{array}{rlll}
      \fl \mbox{(I)} & \displaystyle \a < \a^{*}\equiv{1\over \sqrt{d}+1}, \b & \:\, 
      \displaystyle J_{L} =  {\a(1-\a) \over 1+(d-1)\a} & \:\, 
      \displaystyle \rho_{L} = {\a\over 1+(d-1)\a} \\[13pt]
      \fl \mbox{(II)} & \displaystyle \b < \b^{*}=\a^{*},\, \a \geq \b 
      & \:\, \displaystyle 
      J_{R} = \rho_{R}^{-}= {\b(1-\b)\over 1+(d-1)\b} & \:\,
      \displaystyle \rho_{R}^{+}  = {1-\b\over 1+(d-1)\b}  \\[13pt]
      \fl \mbox{(III)} & \displaystyle (\a,\b) \geq (\a^{*},\b^{*})
      &\:\, \displaystyle  J_{max}= {1 \over (\sqrt{d}+1)^{2}} & \:\,
      \displaystyle \rho_{{N\over 2}} = {1\over \sqrt{d}(\sqrt{d}+1)}.
    \end{array}
    \label{BOUNDARIES1}
  \end{equation} 

Note that for $d=1$ the boundary (and interior) densities predicted by all methods are
equivalent.  However for $d > 1$ we will see that the current and boundary densities in
equations (\ref{BOUNDARIES1}) are significantly more accurate than those predicted by the
simplest mean field theory (Eq. \ref{BOUNDARIES0}).  In fact, the currents and boundary
densities of (\ref{BOUNDARIES1}) match our MC simulation results exactly.  

The qualitative difference between the steady-state boundary densities predicted by
(\ref{BOUNDARIES0}) and (\ref{JL2}), (\ref{JR2}) arise directly from the local, exclusion of the
individual particles.  The boundary densities are accurately predicted with mean field theory;
however, we expect that the predicted density {\it profiles} are as inaccurate as they are for
the standard  $d=1$ TASEP \cite{DER98}.  

The phases (\ref{BOUNDARIES1}) were constructed by compositing results from
approximations unique to each of the three phases. The currents and densities are continuous
at the phase boundaries. In fact, the predicted orders of the phase boundaries, as in the
simple mean field theory,  is consistent with the exact solution of the $d=1$ TASEP
\cite{DER93,DER98,EVANS}. Namely, that $(\partial J/\partial \a)$ and $(\partial J/\partial \b)$
are continuous across the (I) - (III) and (II) - (III) phase boundaries. The phase transition
between (I) and (II) is first order while the transitions between (I) - (III) and (II) - (III) have the
following $d$-dependent discontinuity in the second derivative: 

  \begin{equation}
\fl \quad \mbox{as}\,\, \ve \rightarrow 0, \,\,\,    \displaystyle {\partial^{2} J\over \partial \a^{2}}\bigg|_{\a=\a^{*}+\ve}\!\!\!\!\!-
		  {\partial^{2} J\over \partial \a^{2}}\bigg|_{\a=\a^{*}-\ve} \!\!\!\!\!
		  = {\partial^{2} J\over \partial \b^{2}}\bigg|_{\b=\b^{*}+\ve}\!\!\!\!\! -
		  {\partial^{2} J\over \partial \b^{2}}\bigg|_{\b=\b^{*}-\ve}\!\!\!\!\! = {2 \over \sqrt{d}}.
		  \label{JUMP1}
  \end{equation}

\noindent The nature of the transitions are qualitatively different from those predicted using
the simple mean field approach (\ref{JUMP0}) and  compare favorably with simulations.  

\section{Monte Carlo Simulations}

Extensive Monte Carlo simulations were performed to validate the various analytical models
presented in the previous two sections.  As we expected the densities in the lattice to vary
significantly as the injection and extraction rates were varied, we chose to base our Monte
Carlo code on a continuous-time algorithm developed by Bortz, Kalos, and Lebowitz (BKL)
\cite{BORTZ75}.  The BKL algorithm has the advantage of maintaining a constant
computational efficiency over a wide range of particle densities and has also recently enjoyed
wide use in stochastic simulations of chemical reactions \cite{GILLESPIE}.

The finite-sized effects in our simulations was estimated by running with $d =
1$ and lattices of varying lengths.  For lattices at least one thousand sites long, the MC
results were found to systematically deviate from the known TASEP results by less than half
a percent.  In the simulations with larger particles, the length of the lattice was scaled with
$d$ to ensure that at least one thousand particles could occupy the lattice simultaneously ($N
= 1000d$).  The simulations were run for $4 \times 10^9$ steps, which was sufficient to
reproduce the known TASEP results in lattices much longer than our $d \times 1000$ site
benchmark.  In all simulations, a linear-congruential pseudorandom number generator with a
period of $2 \times 10^{18}$ \cite{NR} was used.

\subsection{Currents}

\begin{figure}
    \begin{center}
      \includegraphics[height=2.5in]{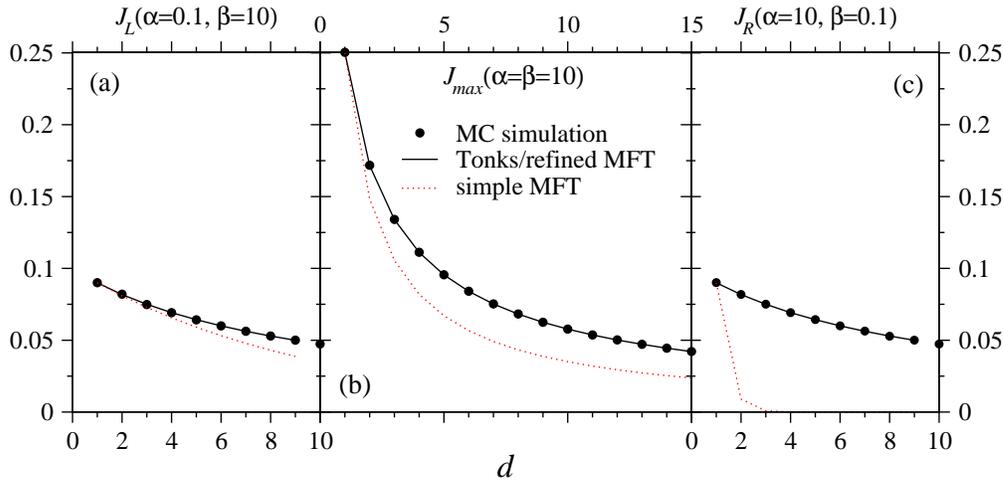}
    \end{center}
    \caption{The currents within each of the phases (I), (II), and (III)
      as a function of integer particle size $d$. Results from 
      simple mean field, refined mean field, and Monte Carlo simulations 
      are shown.
      (a) The current $J_{L}$ as a function of $d$ for $\b = 10$ and 
      $\a=0.1$. (b) The maximal current $J_{max}$ when $\a=\b=10$. 
      (c). The exit limited current $J_{R}$ for $\a=10$ and $\b=0.1$. The simple mean field 
approach always underestimates the currents since it smears out the density and disrupts the 
collective hop of each large particle.}
    \label{CURRENTS}
  \end{figure}

Figure \ref{CURRENTS} plots the steady-state current as a function of particle size $d$ in all
three current regimes. While the simple mean field approximation always underestimates the
currents, the Tonks distribution and the refined mean field (Eqs.  (\ref{JL2}) and (\ref{JR2}))
give currents that match the Monte Carlo results within simulation error ($\approx 1\%$)
 
\begin{figure}
    \begin{center}
      \includegraphics[height=2.6in]{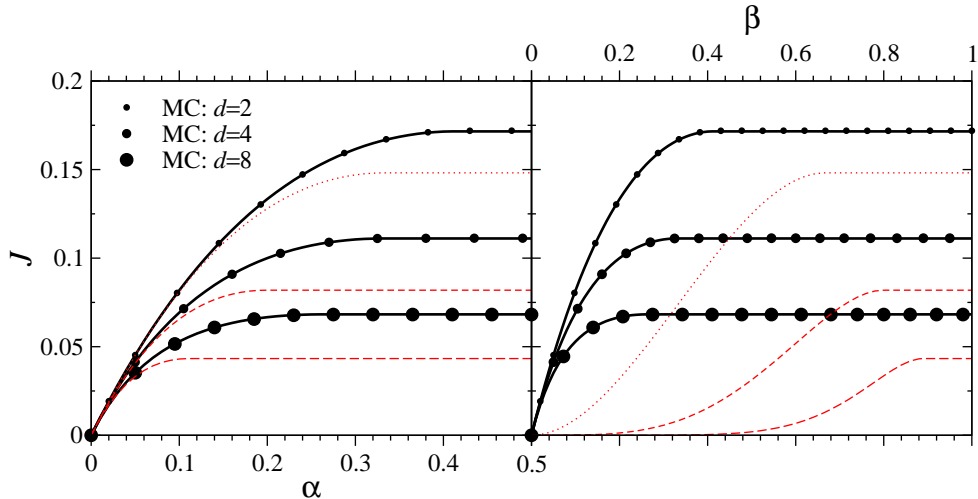}
    \end{center}
    \caption{Currents as functions of $\alpha$ and $\beta$ for fixed $d=2,4,8$.
      (a) Fixed $\b = 10$. (b) Fixed $\a = 10$. The points correspond to 
      MC simulations, the thick solid curves that fit the 
      simulations represent the optimal
      mean field predictions (\ref{BOUNDARIES1}), and 
      the dotted, dashed, long dashed curves are the simple mean field predictions
      for $d=2,4,8$, respectively.}
    \label{CURRENTS2}
  \end{figure}

The currents for various fixed $d$ as functions of $\a$ and $\b$ are shown in Figs. 
\ref{CURRENTS2}.  From these curves one can estimate the critical values $\a^{*}$ and
$\b^{*}$ at which the phase boundaries lie.  The mean field predictions for the currents
displayed in Figs. \ref{CURRENTS2} are a composite of the results from two different
equations.  Equation \ref{JMAX} was used for entrance and exit rates greater than the values
of $\a^{*}, \b^{*}$ in Eq.  (\ref{BOUNDARIES1}). For smaller values of $\alpha,\beta$, Eqs. 
(\ref{JL2}) and (\ref{JR2}) were used. The simulated and refined mean field curves
(\ref{BOUNDARIES1}) are symmetric with respect to the interchange of $\a \leftrightarrow \b$.
However, the simple mean field results (\ref{BOUNDARIES0}) always underestimates the
currents and are shown for comparison by the thin dotted, dashed, and long dashed curves
corresponding to $d=2,4,8$, respectively.  Note the clear asymmetry between entrance and
exit rate limited currents in the simple mean field approximation.

Figures \ref{CURRENTS} and \ref{CURRENTS2} demonstrate the extremely good agreement
between the MC results and our optimal mean-field expressions (\ref{BOUNDARIES1}) for the
current.  In all cases, the difference between the optimal mean field predictions and the MC
results was within error.  The extremely high quality of the fit between the mean field currents
and the MC results strongly supports our conjecture that the currents derived from the Tonks
gas and refined mean field approaches are exact.


\subsection{Particle Densities}

\begin{figure}
    \begin{center}
      \includegraphics[height=3.2in]{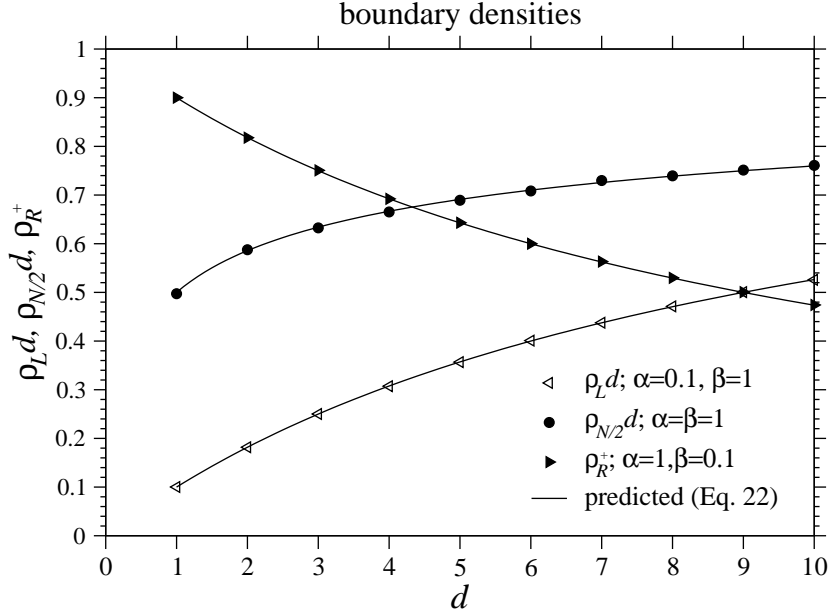}
    \end{center}
    \caption{Representative boundary densities in each of the three current
      regimes. $\triangleleft$ denotes the averaged (over the first $d$ sites) linear
      density $\rho_{1}d$ at the left end when the current is 
      in the entry rate limited
      ($\a=0.1,\b=10$)  low density phase (I). $\bullet$ represents simulation
      results for the averaged density in the lattice interior under maximal current
      conditions $(\a=\b=1)$.  $\blacktriangleright$ represents simulation results
      for the density $\rho_{N}$ at the very last site $N$.
      The thin continuous curves show the predicted densities from 
      the Tonks gas and refined mean field 
      approximations (\ref{JMAX}), (\ref{JL2}), and (\ref{JR2}) and lie within
      simulation noise.}
    \label{BDENSITY2}
  \end{figure}

We now compare the boundary densities predicted in (\ref{BOUNDARIES0}) and
(\ref{BOUNDARIES1}) with those derived from MC simulations.  The MC simulations also
furnish density {\it profiles} which are difficult and tedious to compute using mean field
approximations.  In Figure \ref{BDENSITY2}, the left boundary and interior densities,
multiplied by $d$ ($\rho_{L}d$ and $\rho_{N/2}d$), and the right boundary density $\rho_{R}^{+}$
are shown as functions of particle size $d$.  The net error in each data point was
approximately 3$\%$.  The densities $\rho_{L}d,\rho_{N/2}d$, and $\rho_{R}^{+}$ from our
refined mean field analyses (\ref{BOUNDARIES1}) match the Monte Carlo results
exceptionally well.  

We have also verified that  all  densities, $\rho_{L}, \rho_{N/2}$, and $\rho_{R}^{+}$ can be
readily and accurately computed in all current regimes.  For example, $\rho_{L}$ can be found
from $\a P(x_{1}>d) = J$, where $J=J_{max}$ for the maximal current regime (III), and
$J=J_{R}$ for the exit rate limited regime (II).  Similarly, the interior densities in the entrance
or exit limited regimes are found by equating (\ref{JUNIFORM}) to $J_{L}$ or $J_{R}$ in
(\ref{BOUNDARIES1}) and solving for the appropriate root $\rho$.  The boundary and interior
densities for all regimes are

  \begin{equation}
    \begin{array}{rlcl}
      \fl \: &\qquad  \rho_{L} & \rho_{N/2} &\qquad  \rho_{R}^{+} \\[13pt]
      \fl \mbox{(I)} & \displaystyle {\a \over 1+(d-1)\a} &  {1+(d-1)J_{L}-
	\sqrt{(1+(d-1)J_{L})^2-4dJ_{L}}\over 2d} &  \displaystyle {\a(1-\a)/\b \over 1+(d-1)\a} \\[13pt]
      \fl \mbox{(II)} & \displaystyle {1\over d}-{\b(1-\b)/(\a d)\over 1+(d-1)\b} &  
	  {1+(d-1)J_{R}+\sqrt{(1+(d-1)J_{R})^2-4dJ_{R}}\over 2d}  &  
	  \displaystyle  {1-\b \over 1+(d-1)\b} \\[13pt]
	  \fl \mbox{(III)} & \displaystyle  {(\sqrt{d}+1)^2-1 \over \a d(\sqrt{d}+1)^2} & 
	  \displaystyle {1\over \sqrt{d}(\sqrt{d}+1)} &  \displaystyle {1\over \b (\sqrt{d}+1)^{2}}.
    \end{array}
  \end{equation}

\begin{figure}
    \begin{center}
      \includegraphics[height=4.3in]{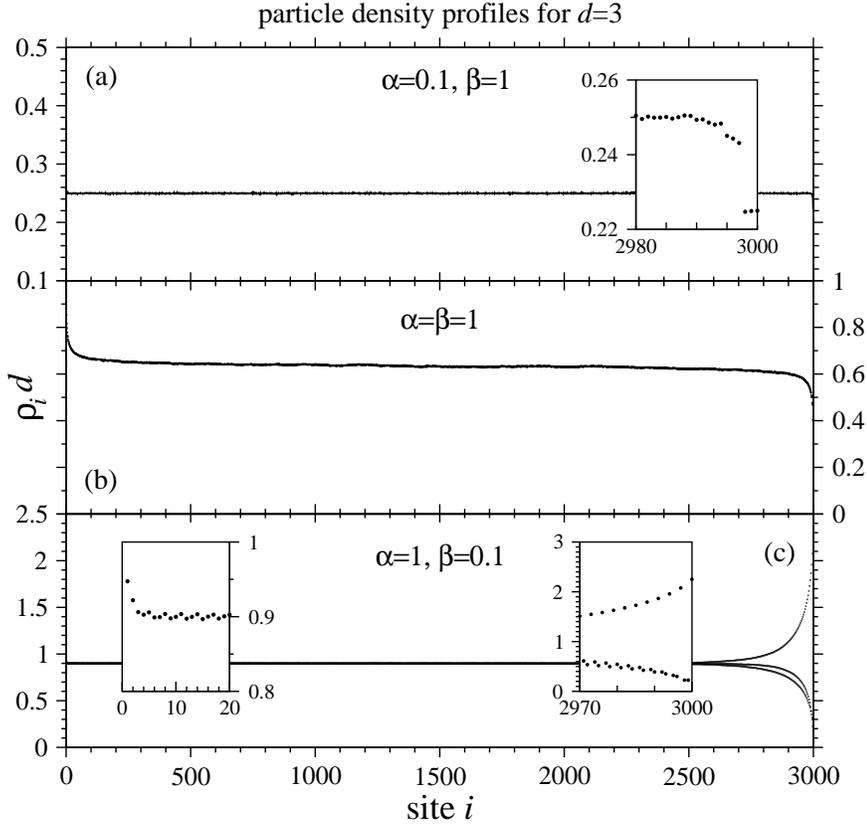}
    \end{center}
    \caption{Normalized density ($\rho_{i}d$) 
      profiles in three representative current regimes. (a) 
      Entrance rate limited regime ($\a=0.1,\b=1$). (b) Maximal current regime ($\a=\b=1$).
      (c) Exit rate limited regime ($\a=1,\b=0.1$). 
      The boundary regions are shown in the insets.}
    \label{PROFILE}
  \end{figure}

 
Recall that in the standard $d=1$ TASEP, mean field theory provides the correct boundary
densities but incorrect density profiles \cite{DER98}.  Figure \ref{PROFILE} depicts the full
filling fractions $\rho d$ as a function of site $i$ for a TASEP with particle size $d=3$. Notice
a that a weak anti-correlation arises at the right boundary even in the entrance rate limited
phase (inset of Fig.  \ref{PROFILE}a). Similarly, there is a very small oscillatory behavior in the
density near the entrance in the exit rate limited regime (first inset of Fig.  \ref{PROFILE}c).  A
more striking anti-correlation is evident near the right boundary of the exit limited chain
(second inset in Fig.  \ref{PROFILE}c) where the density profile splits into approximately two
branches with the branch corresponding to $\rho_{N}, \rho_{N-d}, \rho_{N-2d},\ldots$ having
the highest occupation. This is a consequence of the $d$-site exclusion strongly manifesting
itself near the crowded exit and was used to obtain the refined mean field theory results
(\ref{JR2}).  When currents are exit rate limited, the last particle in the chain spends a
significant amount of time at the last site $i=1000d$. The next site available for
occupation is at $i=999d$ thus forcing $d-1$ sites to be empty as long as the last
particle occupies the last site.  Anti-correlations are negligible at the left boundary since all
particles can only move to the right.  In partially asymmetric exclusion models we expect the
inclusion of backward motion would strengthen the anti-correlation effect at the left boundary
of the lattice, and weaken it at the right.

  \subsection{Phase Boundaries}

  \begin{figure}[h!]
    \begin{center}
      \includegraphics[height=2.9in]{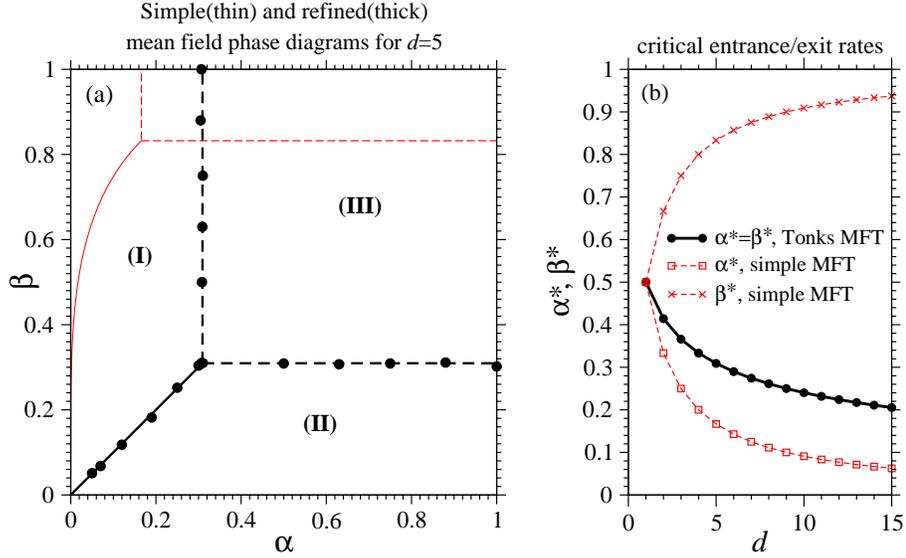}
    \end{center}
    \caption{(a) The phase diagram for the TASEP with particles of size $d=3$ derived from
      the Tonks distribution and the refined mean field (\ref{BOUNDARIES1}). The 
      circles represent the phase boundaries estimated from 
      Monte Carlo simulations.
      For contrast, the $d=3$ phase boundaries derived from the simple mean 
      field model (\ref{BOUNDARIES0}) are 
      overlayed using thin curves and are asymmetric in $\a,\b$. Boundaries across which 
      first order phase transitions occur are shown with solid curves, while 
      those across which second order transitions occur are drawn with 
      dashed segments. (b) The critical values $\a^{*}$ and $\b^{*}$ at which the 
      (I) - (III) and (II) - (III) phase boundaries occur.  For the Tonks gas and refined
      mean field approaches, $\a^{*}=\b^{*}=1/(\sqrt{d}+1)$.
      Critical values derived from the simple mean field (\ref{BOUNDARIES0}) 
      are $\a^{*} = 1/(d+1)$ and $\b^{*} = d/(d+1) \geq 1/2$.}
    \label{PHASE_DIAGRAM}
  \end{figure}

In order to map the phase diagram for the TASEP with finite-sized particles, the boundaries
between the low density, high density, and maximal current phases were determined.  Based
on the good agreement between the MC currents and the currents predicted by equations
(\ref{JL2}) and (\ref{JR2}), we chose to use these equations to predict the location of the
critical values of $\alpha$ and $\beta$.  In the case of the boundary between the maximal
current phase and the entrance/exit rate limited phases, this was accomplished by fitting the
boundary limited current equations to the MC data.  In all cases the fitting parameter was
taken to be $d$, the size of the particles.  The values of $d$ produced by these fits were then
used to find $\a^{*},\b^{*}$ in eqs. (\ref{BOUNDARIES1}).  A variation of this fitting procedure
was employed to determine the location of the boundary between entry and exit limited
phases. Specifically, $J_{L}$ from equation (\ref{JL2})  with the fitted value of $d$ was
equated to a maximal current computed from the MC results.  Equation (\ref{JL2}) was then
solved for $\alpha$ to produce an estimate for the critical value.

  \begin{figure}
    \begin{center}
      \includegraphics[height=2.4in]{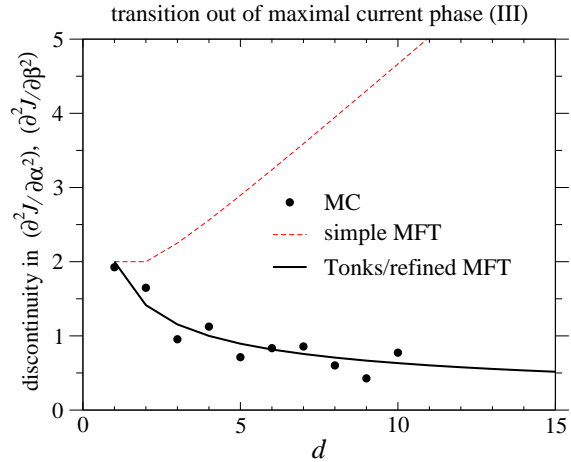}
    \end{center}
    \caption{The discontinuity in the second derivative of the steady-state 
      current with respect to $\a$ or $\b$ at the maximal 
     current phase boundary (I-III and II-III). The simple mean field approach
      (\ref{JUMP0}) 
      predicts an increasing discontinuity with $d$, while 
      the Tonks/refined mean field approach (\ref{JUMP}) 
      predicts a jump that decreases as $2/\sqrt{d}$.}
    \label{JUMP}
  \end{figure}

To determine the order of the phase transitions, numerical derivatives of the current were
computed using a fifth order centred difference formula.  Although the results for the jump in
the second derivatives $\partial^{2}J/\partial \a^{2}$ and $\partial^{2}J/\partial\b^{2}$ have
noise magnified by the differentiation procedure, the results agree quite well (cf. Fig. 
\ref{JUMP}) with the predictions from the Tonks gas/refined mean field approach
(\ref{JUMP1}). An analysis similar to those that have been applied to domain wall
dynamics \cite{KOLO2,POPKOV} is valid in the study of the TASEP with $d > 1$. For example,
additional phase boundaries that delineate regimes of different domain wall dynamics can be
found \cite{KOLO2,POPKOV}. One does not expect that larger ($d>1$) particles to qualitatively change
coarse-grained properties such as shock dynamics. Thus, the additional phase boundaries
separating the regions of different density dynamics can be found by considering shock
velocities of the form $V=(J_{+}-J_{-})/(\rho_{+}-\rho_{-})$ where $J_{\pm}, \rho_{\pm}$ are
the currents and densities to the right or left of a shock \cite{KOLO2,POPKOV}.

\section{Summary and Conclusions}

We have presented mean field theories which accurately model the currents and densities of
a totally asymmetric exclusion process with particles of arbitrary size.  Specifically,
steady-state currents and densities were computed, using both the analytic, mean field
approaches and Monte Carlo simulations. All calculations show that the steady-state currents
qualitatively retain the three phase structure (low density, high density, maximum current)
familiar from the standard ($d=1$) TASEP.
 
Our best analytic approaches utilize an {\it equilibrium} Tonks gas distribution within the maximal current
regime, and an exact enumeration of the most likely occupation states near the lattice ends in the
entrance or exit rate limited regimes.  Unlike the simplest mean field approach, these refined analytic
approximations provide simple formulae for the steady-state currents and densities for all particle
diameters $d$ that agree extremely well (within simulation error) with our Monte Carlo simulation results. 
Based on this close agreement, we conjecture that the  analytic expressions for the currents given by
the Tonks/refined mean field approach (\ref{BOUNDARIES1}) are exact.  The phase diagrams produced
by both of our analytic approaches and the Monte Carlo simulations are shown in Fig. 
\ref{PHASE_DIAGRAM}.  Both analytic approaches predict first order transitions between the high and
low density phases and second order phase transitions between the maximal current and high/low
density phases.  However only the Tonks gas/refined mean field approach produces estimates for the
discontinuities that match the simulation results.  Although mean field methods are not expected to give
correct density profiles near lattice boundaries, a consideration of the diffusion constant $D =
(1/2)(J_{+}-J_{-})/(\rho_{+}-\rho_{-})$, of a coarse-grained density domain wall and the  statistics of the
domain wall position can be used to estimate the  correlation or exponential density decay length $\xi(d)
= \vert \ln(J_{+}(d)/J_{-}(d))\vert$ \cite{KOLO2}.
 
Since mean field results for $d=1$ have been shown to be exact or partially asymmetric models
\cite{RITT,SANDOW}, we expect that our Tonks/refined mean field approach would also provide exact
results for partially asymmetric exclusion processes with $d>1$. One intriguing difference between
$d=1$ and $d>1$ exclusion processes is the nonmonotonic behavior of steady-state densities near the
exit. This behavior arises from the anti-correlations induced by particles of larger size and is a
consequence of the total asymmetry in the dynamics.  Although we have considered only total
asymmetry, analogous expressions can be readily computed for partially asymmetric models by allowing
backward particle steps ($q\neq 0$) and extraction/injection from the left/right. Mean field derivations for
the partially asymmetric case would require the treatment of separate high and low occupancy sites near
both ends of the lattice, as was done for the exit rate limited regime in the TASEP.

The extension of  exclusion processes to larger particles ($d>1$) allows one to directly apply the them
to many biophysical problems including ribosome motion along mRNA and vesicle transport along
microtubules.  Our results, when generalized to include backward hops ($q\neq 0$), may be
useful in determining the optimal scaling of step size and time increments in one-dimensional Brownian
dynamics simulations.  Furthermore, the Tonks gas distribution allows us to 
derive continuum dynamical equations near uniform states:

\begin{equation}
\dot{\rho}(x,t) = (q-p){\partial \rho\over \partial x}{\partial \over \partial x}\left[ {\rho(1-\rho d) \over 1-(d-1)\rho}\right].
\end{equation}

\vspace{3mm}

\noindent {\bf Acknowledgments} TC acknowledges support from the US National Science
Foundation through grant  DMS-0206733.  GL acknowledges support from the Stanford
Graduate Fellowship Program



\end{document}